\documentstyle[preprint,aps,epsf]{revtex}
\begin{document}
\draft
\title{Impurity in a Luttinger liquid: a numerical study
of the finite size energy spectrum and of the orthogonality catastrophe
exponent}

\author{ Shaojin Qin$^\dagger$, Michele Fabrizio$^\star$ and 
Lu Yu$^{\dagger,\ddagger}$}

\address{$^\dagger$ International Center for Theoretical Physics, 
P.O. Box 586, 34100 Trieste, Italy.}

\address{$^\star$ International School for Advanced Studies, Via Beirut 2-4,
I-34014 Trieste, Italy, and Istituto Nazionale di Fisica della Materia, INFM.}

\address{$^\ddagger$ Institute of Theoretical Physics, Academia Sinica,
Beijing 100080, China.}

\date{\today}
\maketitle
\begin{abstract}
The behavior of a single impurity in a one-dimensional Luttinger
liquid is numerically investigated by means of the density matrix
renormalization group. By analyzing the finite size scaling behavior
of the low energy spectrum, we confirm the theoretical prediction
of Kane and Fisher [Phys. Rev. Lett. {\bf 68}, 1220 (1992)] both
for attractive and repulsive interactions. Moreover, we calculate
the exponent of the orthogonality catastrophe, which gives a further
support to the above theoretical prediction.
\end{abstract}

\pacs{71.10.Pm,72.15Nj}

\narrowtext

The problem of an impurity in a one-dimensional (1D) interacting Fermi 
system (Luttinger liquid) has become quite popular in recent years for
its implications to a variety of physical problems as for instance the 
behavior of quantum wires (see e.g. Refs.\onlinecite{Sasha,Exp}) or the 
tunneling through a constriction in the fractional quantum Hall 
regime\cite{Moon}.

The interesting feature of this problem lies in the fact that the
competition between the impurity potential and the electron-electron
interaction leads to quite surprising effects.  If the interaction is 
repulsive, the electrons at low energy see the potential barrier as if 
it were effectively infinite. On the contrary, for an attractive 
interaction the effective scattering close to the Fermi energy is 
vanishingly small (see e.g. Ref.\onlinecite{K&F}).  The two different 
behaviors have a quite simple physical interpretation: in 1D the 
electron gas is quasi-ordered, i.e. it shows power law decaying
correlation functions of the order parameter. Specifically, for 
repulsive interaction it has a charge density wave, while for attractive
interaction a superconductive quasi-long-range order. It is then clear 
that the two different orderings lead to opposite behavior in the 
presence of a local potential, which tends to pin the charge density 
wave but has minor effects for a superconductor.       

>From the theoretical point of view, the problem has many similarities 
with the Kondo effect. In fact the problem of a magnetic impurity 
in a metal can also be transformed into a one-dimensional
model of electrons in the presence of a local potential.
Moreover, in both cases a perturbation expansion in powers of the 
impurity potential breaks down at low energy due to the appearance of 
logarithmic singularities. A standard approach based on renormalization
group leads in both cases to the conclusion that the impurity potential 
flows either to infinity or to zero, depending on the the sign of the 
exchange in the Kondo model or of the interaction in the Luttinger 
liquid. Therefore, analogously to the Kondo 
effect, a low energy description of an impurity in a 
Luttinger liquid needs a correct identification of the low energy fixed 
point.  In particular, Kane and Fisher\cite{K&F} (KF) 
argued that the fixed point corresponds to a chain disconnected at the 
impurity site for repulsive interaction.

This interpretation has been recently questioned by Oreg and
Finkel'stein\cite{Finkelstein} (OF), specifically in the context of the 
X-ray edge singularity. These authors find in fact an exponent of the 
absorption spectra at the threshold which is not equal to the value 
compatible with a disconnected chain fixed point\cite{Xray}. It is 
however unclear whether they would extend their criticism to all the 
results which have been obtained on the basis of the disconnected chain 
fixed point, or if they only doubt about the X-ray edge singularity 
results. In fact, for other properties, like for instance the 
finite size spectrum\cite{A&E} or the conductance through the 
barrier\cite{Moon,Egger}, there exist exact numerical calculations 
which support KF's hypothesis, while for the X-ray edge 
singularity only approximate analytical results are till now 
available\cite{Xray}.

In this paper we investigate this problem by numerical methods, closely 
following the kind of analysis in Ref.\onlinecite{A&E}.
In particular we study the low energy spectrum for 1D periodic chains of 
spinless fermions interacting either with attractive
or repulsive interaction in the presence of a local potential.  
Our numerical results show that this spectrum is compatible
with a disconnected chain fixed point for repulsive interaction and with a
periodic chain if the interaction is attractive.  
We also calculate the exponent of the orthogonality catastrophe 
from the overlap between the ground states in the presence and in the
absence of the local potential. This exponent is related to
the X-ray edge singularity exponent, and therefore our results
can solve the above mentioned controversy. 
We find that the orthogonality catastrophe exponent again 
points in favor of the KF hypothesis.

The development of the density matrix renormalization group (DMRG) 
method\cite{white} has made possible the study of the 
ground state and low excited states of one-dimensional systems with large
number of sites. Previous DMRG studies on $S=1/2$ chains has proved also quite
successful in determining the finite size behavior of the low 
excitation energies, correlations and Friedel's oscillation due to boundary 
effects\cite{ng}.  In 
this paper, we use DMRG to study the low energy spectrum for 
anisotropic $S=1/2$ Heisenberg chains with a local perturbation.
This model can be mapped through a Jordan-Wigner transformation onto a model 
of interacting spinless fermions with a local impurity potential. 

An $S=1/2$ spin-anisotropic Heisenberg chain with open boundary conditions
(OBC) is described by the Hamiltonian
\begin{equation}
H=\sum_{i=1}^{L-1}\left[{J_{xy} \over 2}
                        (S_{i}^+S_{i+1}^-+S_{i}^-S_{i+1}^+)+
                        J_zS_{i}^zS_{i+1}^z \right],
\end{equation}
where $S^+_i$ and $S^-_i$ are spin raising and lowering operators 
at site $i$, $S^z_i$ being its z-component. For periodic boundary 
conditions (PBC) the $L^{th}$ and $1^{st}$ site are also coupled by the 
same exchange couplings $J_{xy}$ and $J_z$.  An impurity site potential is 
introduced by adding a nonzero local magnetic field at the $1^{st}$ site in 
the periodic chain.  An impurity bond is instead introduced by coupling the 
$L^{th}$ and $1^{st}$ site with exchanges $J^{'}_{xy}$ and $J^{'}_z$ 
different from the bulk values $J_{xy}$ and $J_z$. 

Eggert and Affleck\cite{A&E} have shown that the systems flow to the OBC fixed 
point whenever the impurity bond strengths between the $1^{st}$ and 
$L^{th}$ sites differs from the bulk values.  They 
have studied odd length chains under global SU(2) symmetry 
($J_{xy}=J_z$ and $J{'}_{xy}=J{'}_z$) by comparing the parity, 
multiplicity, and magnitude of energies of the ground state and the 
lowest excited states.  We have extended their analysis by considering 
both even and odd length chains,
either for site impurity or bond impurity, and also in the case when 
SU(2) symmetry is explicitly broken.  

Particularly, in the presence of spin anisotropy, we find 
numerically that the low energy spectrum in the presence of a local potential
flows to the OBC spectrum if $0<J_z/J_{xy}<1$, and to the PBC spectrum if 
$-1<J_z/J_{xy}<0$, thus in agreement with the theoretical prediction
of Ref.\onlinecite{K&F}.  For the sake of clarity, we are going to
present our numerical data for the specific case 
of even length chains with bulk anisotropies $J_z/J_{xy}= 0.5$ and 
$-0.5$, which correspond, respectively,  
to repulsive and
attractive interactions in the equivalent spinless fermion 
model. 

We introduce the 
impurity potential by bond couplings $J_z^{'}=b J_{z}$ and 
$J_{xy}^{'}= b J_{xy}$, and denote the anisotropy by $a=J_z/J_{xy}$.  
We calculate the ground state energy $E_0$ and the first excited  
energy $E_1$ for even length chains by DMRG.  The ground state and the 
first excited state are the lowest energy states in the sectors
with $z$-component of the total spin  
$S_z^{tot}=0$ and $1$, respectively. We use the parity of the ground 
state as a reference, and we consider as quantum numbers of the lowest energy 
states their $z$-component of the total spin and their parity. 
Each state is therefore represented by a fixed value of $S_z^{+(-)}$, 
where $+$ is the parity of the ground state and
$-$ denotes the opposite parity. 
The ground state is then represented by $0^{+}$ and the first excited state 
$E_1$ by $1^{-}$.  The two energy levels of the low energy 
spectrum we choose to study are $0^{-}$ and $1^{+}$, and we denote 
their energies by $E_2$ and $E_3$, respectively.  

For different impurity bond strength $b$, 
$L[E_1(a,b)-E_0(a,b)]=\pi v_F $ is a bulk property and therefore 
remains the same for fixed anisotropy $a$.
This is shown in Fig.1 for $a=\pm 0.5$ and $b=0,0.1$ and $1$, 
($b=0$ and $1$ are the values for OBC and PBC, respectively).  
The behavior of the low energy spectrum in the presence of the local
potential is instead made evident by the flow of the two 
energy levels $E_2$ and $E_3$.  We use the energy unit $\pi v_F/L$, 
i.e. we scale the energy as
\begin{equation}
\begin{array}{l}
e_2(a,b)= \displaystyle \frac{E_2(a,b)-E_0(a,b)}{E_1(a,b)-E_0(a,b)},\\
e_3(a,b)= \displaystyle \frac{E_3(a,b)-E_0(a,b)}{E_1(a,b)-E_0(a,b)}.
\end{array}
\end{equation}

For attractive electron-electron interaction ($a=-0.5$), we plot in 
Fig.2 $e_2(-0.5,b)$ and $e_3(-0.5,b)$ for different $b$-values:
$b=0$, corresponding to OBC, $b=1$ for PBC, and 
$b=0.1$ for an intermediate case.  It is clear from Fig.2 that 
$e_2(-0.5,0.1)$ and $e_3(-0.5,0.1)$ flow to $e_2(-0.5,1)$ and $e_3(-0.5,1)$, 
respectively, i.e.
they flow to PBC. We just mention that other 
energy levels of the spectrum for attractive interaction 
$-1<a<0$ also flow to those with PBC.  
For repulsive electron-electron interaction ($a=0.5$), we plot in 
Fig.3 $e_2(0.5,b)$ and $e_3(0.5,b)$ for different $b$-values.
The energy levels of the spectrum for repulsive interaction 
flow to those with OBC.  

The same conclusions can be drawn for other values of the
spin-anisotropy $a$ and other impurity bond strength $b$,
or by substituting the bond potential with a site potential
provided by a local magnetic field, or else for the odd length
chain case (which however has already been extensively discussed in 
Ref.\onlinecite{A&E}).

Therefore our numerical results show that PBC is the fixed point 
towards which the system in the presence of a local potential flows
when the interaction is attractive, while 
OBC is the fixed point for systems with repulsive 
interaction.

An interesting quantity which can also be evaluated numerically and which
further supports the interpretation of the strong coupling fixed point
in terms of a perfectly reflecting barrier is the exponent $\alpha$ of
the orthogonality catastrophe\cite{Anderson}.  This exponent is defined 
through the overlap between the ground state wave functions for a 
system of size $L$ in the presence, $|\phi\rangle$, and in the absence, 
$|\phi_0\rangle$, of the impurity potential. In particular,
\begin{equation}
\langle \phi | \phi_0 \rangle \sim \left(\frac{1}{L}\right)^\alpha.
\label{overlap}
\end{equation}
The orthogonality exponent is intimately related to the 
exponent of the X-ray edge singularity. Let us assume that the 
scattering potential is present if some localized level is empty and is 
absent otherwise. The localized state Green function can be shown to 
decay at long times like:
\begin{equation}
\begin{array}{lcl}
G(t) & = & \langle\phi_0 |  d^\dagger(t) d(0) |\phi_0  \rangle \\
     & = & \langle\phi_0 | {\rm e}^{i\hat{H}_0 t}{\rm e}^{-i\hat{H} t} 
|\phi_0 \rangle \sim 
{\rm e}^{- i E_{edge}t} \left( \frac{1}{t} \right)^{2\alpha},
\end{array}
\label{G}
\end{equation}
where $d$ is the operator which empties the localized level,
$\hat{H}_0$ and $\hat{H}$ the Hamiltonians in the absence and in the
presence of the scattering potential, respectively.  In the absence of 
electron-electron interaction, the exponent $\alpha$ can be exactly 
determined\cite{Anderson} and is given by:
\begin{equation}
2 \alpha = \left( \frac{\delta_e}{\pi} \right)^2 +
 \left( \frac{\delta_o}{\pi} \right)^2.
\end{equation}
The phase shifts of the even ($\delta_e$) and odd ($\delta_o$) scattering 
channels can in turn be related to the properties of the barrier, since
\begin{equation}
\delta_{e(o)} = \frac{1}{2} \left( \delta_+ \pm \tan^{-1} \frac{|r|}{|t|}
\right),
\label{delta}
\end{equation} 
where the transmission amplitude $t=|t|{\rm e}^{i\delta_+}$, and $r$ is 
the reflection coefficient.

In order to simplify the interpretation of the numerical results, we 
consider an anisotropic Heisenberg chain of length $L$. The ground 
state is in the subspace with total spin $z$-component $S_z=0$. The 
impurity potential is provided by modifying the exchange of a single 
bond in the chain. This model maps onto a chain of interacting spinless 
fermions at half filling. The 
impurity potential is such as to preserve the particle-hole symmetry 
of the Hamiltonian.  As a consequence, the forward scattering phase 
shift $\delta_+= 0$, since a finite $\delta_+$ would imply a breaking 
of the particle-hole symmetry. Before presenting the numerical results, 
it is worthwhile to discuss what the theory would predict for the 
exponent $\alpha$.  As we already mentioned, according to Kane and 
Fisher the strong coupling fixed point towards which the model flows in 
the long wavelength limit consists of an Heisenberg open chain: the 
impurity has simply changed the boundary conditions from periodic to 
open. In this case the exponent is predicted to be  
$\alpha = 1/16$\cite{Xray}, which corresponds simply to 
take $\delta_+=0$ and $|r/t|\to \infty$ in Eq.(\ref{delta}).
As we previously said, the KF 
interpretation has been questioned by OF 
in the context of the X-ray edge singularity\cite{Finkelstein}. 
Following them, we expand the Green function (\ref{G}) in powers of the 
impurity potential.  Each term in the perturbation expansion can be 
evaluated exactly in the long-time limit and finally we find, in 
accordance with OF, that     
the imaginary time Green function $G(\tau)$ coincides with the partition 
function of a two-dimensional classical Coulomb gas confined on a line of 
length $\tau$. The relevance of the impurity potential translates in the 
language of the Coulomb gas into an increasing fugacity: the gas is 
therefore in the plasma phase. If $\tau$ is much bigger than the 
screening length of the plasma $\tau_{scr}$, then OF have predicted that
\begin{equation}
\lim_{\tau\gg \tau_{scr}} G(\tau) = {\rm e}^{-E_{edge}\tau} const.,
\label{Z-tau}
\end{equation}
which would imply an exponent $\alpha=0$. 
As we see the two theoretical
approaches lead to two completely different results; therefore the 
exact numerical approach we are going to describe turns out to be quite 
decisive.    

We denote the overlap integrals to be evaluated numerically and their 
exponents as:
\begin{equation}
\begin{array}{l}
O_{P}(L)=\langle \phi_{PBC} |\phi\rangle \propto (\frac{1}{L})^\alpha,\\
O_{O}(L)=\langle \phi_{OBC} |\phi\rangle \propto (\frac{1}{L})^\beta,
\end{array}
\end{equation}
where the ground states are $|\phi_{PBC}\rangle$ for PBC,
$|\phi_{OBC}\rangle$ for OBC, and $|\phi\rangle$ for chains
with PBC but in the presence of a modified bond $0<b<1$. 
The exponents $\alpha$ and $\beta$ are given by
\begin{equation}
\begin{array}{l}
\alpha(a,b)= \displaystyle \frac{\ln O_P(L+2)- \ln O_P(L)}
                                {\ln L - \ln (L+2)},\\
\beta(a,b)=  \displaystyle \frac{\ln O_O(L+2)- \ln O_O(L)}
                                {\ln L - \ln (L+2)}.
\end{array}
\end{equation}
For anisotropy $a=\pm 0.5$ and impurity bond strength $b=0.1$, $0.9$, we
calculate the ground states by DMRG method, and then calculate the 
exponents $\alpha(a,b)$ and $\beta(a,b)$.  We plot in Fig.4 the 
exponents as a function of $-2/[\ln L +\ln (L+2)]$ for different
sizes $L$. The figure shows that 
$\alpha(a,b)$ flows to $1/16$ and $\beta(a,b)$ to $0$ for repulsive 
interaction ($a=0.5$), and vice versa for attractive interaction 
($a=-0.5$).

The values $\alpha=0$ and $\beta=1/16$ which we find for attractive 
interaction are in agreement with the flowing of the system to the PBC 
fixed point.  On the other hand, $\alpha=1/16$ 
and $\beta=0$ for repulsive interaction show that the 
system flows indeed to the OBC fixed point, thus in agreement
with Kane and Fisher's
interpretation of the strong coupling fixed point and with 
the analytical results of Refs.\onlinecite{Xray}.

In conclusion, we have investigated by numerical DMRG method
the impurity problem in a Luttinger liquid. All our results,
which include a detailed analysis of the finite size scaling
behavior of the low energy spectrum and the evaluation of
the orthogonality catastrophe exponent, suggest that at
low energy the barrier acts as it were perfectly reflecting.
The orthogonality catastrophe exponent result may help to clarify
the controversy recently raised by the authors of 
Ref.\onlinecite{Finkelstein} about the X-ray edge 
singularity\cite{comment}.    
 
L.Y. would like to thank I. Affleck
for helpful discussions and communications. Mobility in Europe
involved in this research project was partly sponsored by EEC
under contract ERB CHR XCT 940438.

\begin{figure}[hbt]
\caption{
The first excitation energy $E_1(a,b)-E_0(a,b)$ multiplied by chain 
length $L$ is plotted v.s. $1/\ln L$, $L=4,6,\cdots,60$, for different
values of the
anisotropy $a=\pm 0.5$ and impurity bond strength $b=0,1,0.1$. 
}
\end{figure}

\begin{figure}[hbt]
\caption{
The scaled low energy levels $e_2(a,b)$ and $e_3(a,b)$ of Eq.(2) for 
attractive interaction ($a=-0.5$) are plotted v.s. $1/\ln L$,  
$L=4,6,\cdots,50$.  $e_2(-0.5,0)$, $e_3(-0.5,0)$ is for OBC ($b=0$), 
$e_2(-0.5,1)$, $e_3(-0.5,1)$ for PBC ($b=1$), and $e_2(-0.5,0.1)$, 
$e_3(-0.5,0.1)$ for impurity chains. 
}
\end{figure}
 
\begin{figure}[hbt]
\caption{
The scaled low energy levels $e_2(a,b)$ and $e_3(a,b)$ of Eq.(2) for 
repulsive interaction ($a=0.5$) is plotted v.s. $1/\ln L$,  
$L=4,6,\cdots,40$.  $e_2(0.5,0)$, $e_3(0.5,0)$ is for OBC ($b=0$), 
$e_2(0.5,1)$, $e_3(0.5,1)$ for PBC ($b=1$), and $e_2(0.5,0.1)$, 
$e_3(0.5,0.1)$ for impurity chains.  
}
\end{figure}
 
\begin{figure}[hbt]
\caption{
The exponents $\alpha(a,b)$ and $\beta(a,b)$ of Eq.(10) are plotted for 
both repulsive and attractive interactions $a=\pm 0.5$ 
and impurity bond strength $b=0.1,0.9$ as a function of 
$-2/[\ln L +\ln (L+2)]$ with $L=4,6,\cdots,48$.  The fitting 
lines indicate that $\alpha(-0.5,0.1)$ and the $\beta(0.5,0.1)$ 
flow to zero, while $\alpha(0.5,0.1)$ and $\beta(-0.5,0.1)$ 
flow to $1/16$. 
}
\end{figure}
 

\begin{references}
\bibitem{Sasha}A.O. Gogolin, Ann. Phys. (Paris) {\bf 19}, 411 (1994)
\bibitem{Exp}S. Tarucha, T. Honda and T. Saku,
Sol. State Comm. {\bf 94}, 413 (1995).
\bibitem{Moon} K. Moon, C.L. Kane, S.M. Girvin and M.P.A. Fisher,
Phys. Rev. Lett. {\bf 71}, 4381 (1993).
\bibitem{K&F}C.L. Kane and M.P.A. Fisher, Phys. Rev. Lett. {\bf 68},
1220 (1992).
\bibitem{A&E} S. Eggert and I. Affleck, Phys. Rev. B {\bf 46},
10866 (1992).
\bibitem{Egger} K. Leung, R. Egger, and C.H. Mak, Phys. Rev. Lett.
{\bf 75}, 3344 (1995).
\bibitem{Finkelstein} Y. Oreg and A.M. Finkel'stein, Phys. Rev. B
{\bf 53}, 10928 (1996).  
\bibitem{Xray}A.O. Gogolin, Phys. Rev. Lett. {\bf 71}, 2995 (1993);
N.V. Prokof'ev, Phys. Rev. B {\bf 49}, 2148 (1994);
C.L. Kane, K.A. Matveev, and L.I. Glazman, {\sl ibid.} {\bf 49}, 2253 (1994), 
I. Affleck and W.W. Ludwig, J. Phys. A {\bf 27}, 5375 (1994).
\bibitem{white} S.R.\ White, \prl {\bf 69}, 2863 (1992), 
     \prb {\bf 48}, 10345 (1993).
\bibitem{ng} T.K.\ Ng, S.\ Qin and Z-B.\ Su, \prb (to be pblished, BM5930);
             {\sl ibid.} {\bf 52}, 12844 (1995).
\bibitem{Anderson}P.W. Anderson, Phys. Rev. Lett. {\bf 18}, 1049 (1967).
\bibitem{comment}Unless the half filling is a special situation [
A.M. Finkel'stein, private communication].
\end{references}
\end{document}